\documentclass[aps,superscriptaddress,showpacs]{revtex4}
\usepackage{amssymb}


\begin{document}
\thispagestyle{empty}

\title{Reanalysis of the $e^+e^-\to\eta\gamma$ reaction cross section}

\begin{abstract}
In the experiment with the SND detector at the VEPP-2M  $e^+e^-$ collider
measuring the $e^+e^-\to \eta\gamma$ cross section in the energy range
$0.6<\sqrt{s}<1.38\,\mathrm{GeV}$ the reanalysis of data is performed.
The goal is to improve the accuracy of the previous results
by analysing ambiguities in the approximation of energy dependence
of the $e^+e^-\to \eta\gamma$ cross section, which were not taken into
account in our previous study. We report new results
on the approximation of the $e^+e^-\to \eta\gamma$
cross section based on Vector Dominance Model
under new model parameter assumptions. 
\end{abstract}


\author{M.~N.~Achasov}
\affiliation{Budker Institute of Nuclear Physics, Novosibirsk, 630090, Russia}
\affiliation{Novosibirsk State University, Novosibirsk, 630090, Russia}

\author{K.~I.~Beloborodov}
\affiliation{Budker Institute of Nuclear Physics, Novosibirsk, 630090, Russia}
\affiliation{Novosibirsk State University, Novosibirsk, 630090, Russia}

\author{A.V.~Berdyugin}
\affiliation{Budker Institute of Nuclear Physics, Novosibirsk, 630090, Russia}

\author{A.~G.~Bogdanchikov}
\affiliation{Budker Institute of Nuclear Physics, Novosibirsk, 630090, Russia}

\author{A.~D.~Bukin}
\affiliation{Budker Institute of Nuclear Physics, Novosibirsk, 630090, Russia}
\affiliation{Novosibirsk State University, Novosibirsk, 630090, Russia}

\author{D.~A.~Bukin}
\affiliation{Budker Institute of Nuclear Physics, Novosibirsk, 630090, Russia}

\author{T.~V.~Dimova}
\affiliation{Budker Institute of Nuclear Physics, Novosibirsk, 630090, Russia}

\author{V.~P.~Druzhinin}
\affiliation{Budker Institute of Nuclear Physics, Novosibirsk, 630090, Russia}
\affiliation{Novosibirsk State University, Novosibirsk, 630090, Russia}

\author{V.~B.~Golubev}
\affiliation{Budker Institute of Nuclear Physics, Novosibirsk, 630090, Russia}
\affiliation{Novosibirsk State University, Novosibirsk, 630090, Russia}

\author{A.~A.~Korol}
\affiliation{Budker Institute of Nuclear Physics, Novosibirsk, 630090, Russia}

\author{S.~V.~Koshuba}
\affiliation{Budker Institute of Nuclear Physics, Novosibirsk, 630090, Russia}

\author{E.~V.~Pakhtusova}
\affiliation{Budker Institute of Nuclear Physics, Novosibirsk, 630090, Russia}

\author{S.~I.~Serednyakov}
\affiliation{Budker Institute of Nuclear Physics, Novosibirsk, 630090, Russia}
\affiliation{Novosibirsk State University, Novosibirsk, 630090, Russia}

\author{Z.~K.~Silagadze}
\affiliation{Budker Institute of Nuclear Physics, Novosibirsk, 630090, Russia}
\affiliation{Novosibirsk State University, Novosibirsk, 630090, Russia}

\author{A.~V.~Vasiljev}
\affiliation{Budker Institute of Nuclear Physics, Novosibirsk, 630090, Russia}
\affiliation{Novosibirsk State University, Novosibirsk, 630090, Russia}


\maketitle

In our previous work \cite{snd_etag} we performed a measurement of the
$e^+e^-\to \eta\gamma$ reaction cross section for two $\eta $-meson
decay modes: $\eta\to 3\pi^0$ and $\eta \to \pi^+\pi^-\pi^0$. The measured
cross section was approximated within Vector Dominance Model (VDM). In this 
approximation the fact that with the free magnitudes and phases of all four
$\rho$, $\omega$, $\phi$, and $\rho^{\prime}$ resonances, the minimized likelihood
function has several minima and thus, a choice of the right
minimization solution is ambiguous, was ignored.  Additional physical restrictions
must be applied to, at least, the parameters of the $\rho^{\prime}$
resonance, in order to make a right choice of the minimization solution.

In this work the $\rho^{\prime}$-meson parameters were estimated from the
experimental data on the $e^+e^-\to\eta\rho$ cross section \cite{cmdetarho,dm2etarho}.
The cross section value in the $\rho^\prime$-resonance maximum
obtained from the approximation of the $e^+e^-\to\eta\rho$ cross section 
can be translated to the $e^+e^-\to \rho^{\prime} \to \eta\gamma$ cross
section using VDM as in Ref.~\cite{cmdetagamma}.
The approximation of
the data on the $e^+e^-\to \eta\rho$ cross section was done with three different
models describing the energy dependence of the $\rho^{\prime}$ resonance width:
1 --- $B(\rho^{\prime}\to\omega\pi)=0.5$, $B(\rho^{\prime}\to\pi^+\pi^-)=0.5$;
2 --- $B(\rho^{\prime}\to\omega\pi)=1$;
and 3 --- $B(\rho^{\prime}\to \pi^+\pi^-)=1$. Under these assumptions the 
resulting mass,
width, and maximum cross section of the $\rho^{\prime}$ resonance varied
within 1440--1520~MeV, 220--400~MeV, and 0.08--0.13~nb, respectively.
All three $\rho^{\prime}$ parameter sets were used in the
$e^+e^-\to\eta\gamma$ cross section
approximation. The difference in the experimental results was included in
the systematic uncertainty of the final result. The cross section at the
$\rho^{\prime}$ resonance maximum acquired an additional 20\% systematic error
from the uncertainty of the cross section calculation in the VDM.

Natural values for the phase of the $\rho^{\prime}$ amplitude are 0 or
180 degrees. The phase shift from these values can be induced by the
$\rho$ and $\rho^{\prime}$ mixing via common decay channels, for example,
$\rho,\rho^{\prime}\to \omega\pi^0$. In the model considered in 
Ref.~\cite{achasov-rho} it was shown that mixing is strongly energy
dependent. For the approximation of the data on $e^+e^-\to \eta\gamma$
cross section the $\varphi_{\rho^{\prime}}$ value at the $\phi$
resonance maximum is important.
Since at this energy the $\rho-\rho^{\prime}$ mixing is small,
no significant phase shifts from 0 or 180 degrees are
expected. Two approximations with the $\varphi_{\rho^{\prime}}$ phase fixed
at these values were considered. In order to estimate a corresponding
systematic uncertainty of the approximation parameters due to the 
$\rho^{\prime}$ phase uncertainty, we varied $\varphi_{\rho^{\prime}}$
by $\pm20^\circ$ with respect to its initial values.

The fitting results with the  $\rho^{\prime}$ parameters fixed to described above
values are presented in Table~\ref{tabresfit}. Excluded from Table~\ref{tabresfit} 
are two solutions with the
phase $\varphi_\omega \sim 180^\circ$. The values of 
$B(\omega\to\eta\gamma)\simeq (28\pm 1)\times 10^{-4}$ corresponding to these
values of  $\varphi_\omega$  contradict the result of Ref. \cite{GAMS}
$B(\omega\to\eta\gamma)= (8.3\pm 2.1)\times 10^{-4}$, where the contribution from
$\rho-\omega$ interference was suppressed by event selection cuts.
The quark model predictions for the phases of $\omega$ and $\phi$
mesons are $0^\circ$ and $180^\circ$,
respectively. The values close to those were obtained, for example, for 
the processes $e^+e^-\to 3\pi$ \cite{SND_3pi} and $e^+e^-\to \pi^0\gamma$ \cite{SND_pi0g}.
The difference between the observed phases and quark model predictions of
$0^\circ$ and $180^\circ$
for these processes can be described by $\phi-\omega$ and $\rho-\omega$ mixing \cite{achasov_3pi,SND_3pi}.
Estimations show, that the contribution of the $\phi-\omega$ mixing in the $\eta\gamma$
channel is small. Taking into account that $\phi-\rho$ mixing is small
($B(\phi\to\pi^+\pi^-)\sim 10^{-4}$), the expected deviation of the $\phi$-meson
phase from $180^\circ$ does not exceed few degrees. The $\rho-\omega$ mixing leads to the
$\omega$-meson phase $\varphi_\omega\sim 15^\circ$. Since for the solution corresponding
to the $\varphi_{\rho^\prime}=180^\circ$ the value of $\varphi_\phi$ differs
from its expectation by more than $3.5\sigma$, for our final result we 
take the solution corresponding to $\phi_{\rho^\prime}=0^\circ$. 
It should be noted that the same choice of the
$\rho^\prime$ phase was used in the work \cite{cmdetagamma}.

\begin{table}
\caption{\label{tabresfit} Results of the approximation. The first error is
statistical, the second is systematic.}
\begin{ruledtabular}
\begin{tabular}[t]{ccc}
$\varphi_{\rho^\prime}$          & $0^{\circ}$                       & $180^{\circ}$                    \\ \hline
$\sigma_{\rho\to\eta\gamma}$     & $(0.322 \pm 0.034 \pm 0.019)$~nb  & $(0.319 \pm 0.032 \pm 0.019)$~nb \\
$\sigma_{\omega\to\eta\gamma}$   & $(0.744 \pm 0.075 \pm 0.027)$~nb  & $(0.816 \pm 0.081 \pm 0.029)$~nb \\
$\sigma_{\phi\to\eta\gamma}$     & $(57.21 \pm 0.95 \pm 1.66)$~nb    & $(59.79 \pm 0.75 \pm 1.73)$~nb   \\
$\varphi_{\omega}$               & $(6.9 \pm 7.7 \pm 2.3)^{\circ}$   & $(18.3 \pm 7.7 \pm 2.4)^{\circ}$ \\
$\varphi_{\phi}$                 & $(166 \pm 18 \pm 8)^{\circ}$      & $(219 \pm 7 \pm 9)^{\circ}$      \\ 
\end{tabular}
\end{ruledtabular}
\end{table}

Using the world-average values \cite{pdg} for the $\rho$, $\omega$, and $\phi$
masses and $\rho,\omega,\phi \to e^+e^-$ decay probabilities, we can calculate
the following values for the decay probabilities into the $\eta\gamma$ final state:

\begin{eqnarray}
\label{bretg}
B(\rho\to\eta\gamma)   & = & (2.82 \pm 0.30 \pm 0.17)\times 10^{-4},    \nonumber \\
B(\omega\to\eta\gamma) & = & (4.33 \pm 0.44 \pm 0.17)\times 10^{-4},              \\
B(\phi\to\eta\gamma)   & = & (1.364 \pm 0.023 \pm 0.044)\times 10^{-2}. \nonumber
\end{eqnarray}

The systematic error on the branching fractions and cross sections quoted in 
Table~\ref{tabresfit}
and Eq.~(\ref{bretg}) includes a 1.9\% uncertainty of the detection efficiency, 2\% uncertainty
of the integrated luminosity, and the model uncertainty equal to 5.2\% for the
$\sigma_{\rho\to\eta\gamma}$, 2.3\% for the $\sigma_{\omega\to\eta\gamma}$, and
0.9\% for the $\sigma_{\phi\to\eta\gamma}$. The systematic uncertainty of the
decay probability also includes the error on the $\rho,\omega,\phi \to e^+e^-$
branching fractions.

Using the results of the cross section approximation we calculated the radiative
correction and its model uncertainty. The model uncertainty is included into
a systematic error on the total cross section (Tables~\ref{tabcrs1} and \ref{tabcrs2}) 
together with
the errors on detection efficiency and integrated luminosity. The model error
does not exceed 1\% for all energy points below 1040~MeV. At the energy of
about 1100~MeV near the cross section minimum it reaches a 17\% level.

In comparison with the results published in Ref. \cite{snd_etag}, $B(\rho\to \eta\gamma)$
increased by $1.4\sigma$, $B(\omega\to \eta\gamma)$ decreased by $0.7\sigma$, the
change in $B(\phi\to \eta\gamma)$ is less than $0.1\sigma$. The alterations 
of the measured cross section values are statistically insignificant in the energy range 
below 1030~MeV. For energies in the range 1030--1050~MeV and above 1200~MeV the
difference is about one statistical standard deviation, but in the
energy range 1050-1200~MeV, where the reaction cross section is small, the difference
between the cross sections corresponds to six statistical standard deviations.

\section{Conclusion}
In Table~\ref{TabBrCompar} the decay probabilities presented in our previous
paper \cite{snd_etag}, the results of this work, and other most precise experiments
are listed. The results on decay branching fractions obtained in this work
differ from \cite{snd_etag} by about one standard deviation, which proves the
stability of the results under different model assumptions.

\begin{table}
\caption{\label{tabcrs1} Cross section ($\sigma$) of the process
$e^+e^-\to\eta\gamma$ measured in the decay mode
$\eta\to 3\pi^0$. $E$ is center-of-mass energy.
The first error is statistical, the second --- systematic.}
\begin{ruledtabular}
\begin{tabular}[t]{cccccc} 
 $E$, MeV & $\sigma$, nb                        &  $E$, MeV & $\sigma$, nb                    &  $E$, MeV & $\sigma$, nb                        \\ \hline
 600.00   & $< 0.36~~~90\% CL$                  &  794.24   & $0.58^{+0.17}_{-0.13} \pm 0.02$ & 1019.62   & $54.14 \pm 0.99          \pm 1.52$  \\
 630.00   & $< 0.19~~~90\% CL$                  &  800.29   & $0.56^{+0.14}_{-0.11} \pm 0.02$ & 1020.58   & $41.37 \pm 1.60          \pm 1.16$  \\
 660.00   & $0.059^{+0.078}_{-0.038} \pm 0.002$ &  810.26   & $0.29^{+0.11}_{-0.08} \pm 0.01$ & 1021.64   & $23.37 \pm 1.16          \pm 0.66$  \\
 690.00   & $< 0.10~~~90\% CL$                  &  820.00   & $0.45^{+0.12}_{-0.09} \pm 0.01$ & 1022.78   & $12.30 \pm 0.66          \pm 0.35$  \\
 720.00   & $0.22^{+0.07}_{-0.05}    \pm 0.01$  &  840.00   & $0.25^{+0.06}_{-0.05} \pm 0.01$ & 1027.76   & $ 2.84 \pm 0.19          \pm 0.08$  \\
 750.26   & $0.33^{+0.14}_{-0.10}    \pm 0.01$  &  880.00   & $0.19^{+0.08}_{-0.06} \pm 0.01$ & 1033.70   & $ 0.86 \pm 0.09          \pm 0.03$  \\
 760.29   & $0.36^{+0.14}_{-0.10}    \pm 0.01$  &  920.00   & $0.27^{+0.08}_{-0.06} \pm 0.01$ & 1039.68   & $ 0.34 \pm 0.04          \pm 0.01$  \\
 764.31   & $0.32^{+0.13}_{-0.09}    \pm 0.01$  &  940.00   & $0.27^{+0.08}_{-0.06} \pm 0.01$ & 1049.76   & $ 0.12 \pm 0.02          \pm 0.01$  \\
 770.31   & $0.54^{+0.15}_{-0.12}    \pm 0.02$  &  950.00   & $0.16^{+0.09}_{-0.06} \pm 0.01$ & 1059.76   & $ 0.072 \pm 0.011        \pm 0.005$ \\
 774.23   & $0.77^{+0.20}_{-0.16}    \pm 0.02$  &  958.00   & $0.39^{+0.14}_{-0.11} \pm 0.01$ & 1078.54   & $0.032^{+0.009}_{-0.007} \pm 0.003$ \\
 778.21   & $1.49^{+0.25}_{-0.22}    \pm 0.04$  &  970.00   & $0.23^{+0.11}_{-0.08} \pm 0.01$ & 1099.92   & $0.016^{+0.007}_{-0.005} \pm 0.001$ \\
 780.14   & $1.56^{+0.23}_{-0.20}    \pm 0.04$  &  980.00   & $0.24^{+0.19}_{-0.12} \pm 0.01$ & 1131.58   & $0.020^{+0.009}_{-0.007} \pm 0.003$ \\
 781.16   & $2.17^{+0.25}_{-0.23}    \pm 0.06$  &  984.10   & $0.42^{+0.12}_{-0.09} \pm 0.01$ & 1182.96   & $0.020^{+0.016}_{-0.010} \pm 0.003$ \\
 782.06   & $2.11^{+0.17}_{-0.16}    \pm 0.06$  & 1003.82   & $ 1.61 \pm 0.19       \pm 0.05$ & 1227.34   & $0.015^{+0.035}_{-0.012} \pm 0.001$ \\
 783.17   & $1.98^{+0.20}_{-0.18}    \pm 0.06$  & 1009.68   & $ 3.20 \pm 0.30       \pm 0.09$ & 1271.68   & $0.051^{+0.049}_{-0.028} \pm 0.001$ \\
 784.27   & $1.82^{+0.22}_{-0.12}    \pm 0.05$  & 1015.64   & $15.15 \pm 0.86       \pm 0.42$ & 1315.44   & $0.021^{+0.048}_{-0.017} \pm 0.001$ \\
 785.40   & $1.50^{+0.26}_{-0.23}    \pm 0.04$  & 1016.70   & $25.06 \pm 1.22       \pm 0.70$ & 1360.44   & $0.022^{+0.029}_{-0.014} \pm 0.001$ \\
 786.21   & $1.30^{+0.22}_{-0.19}    \pm 0.04$  & 1017.66   & $35.55 \pm 1.73       \pm 1.00$ &           &                                     \\
 790.22   & $0.83^{+0.21}_{-0.17}    \pm 0.02$  & 1018.64   & $53.75 \pm 1.54       \pm 1.51$ &           &                                     \\
\end{tabular}
\end{ruledtabular}
\end{table}

\begin{table}
\caption{\label{tabcrs2} Cross section ($\sigma$) of the process
$e^+e^-\to\eta\gamma$ measured in the decay mode
$\eta\to\pi^+\pi^-\pi^0$. $E$ is center-of-mass energy.
The first error is statistical, the second --- systematic.}
\begin{ruledtabular}
\begin{tabular}[t]{cccccc} 
 $E$, MeV & $\sigma$, nb              & $E$, MeV  & $\sigma$, nb              & $E$, MeV  & $\sigma$, nb                 \\ \hline
 755.54   & $ 0.19 \pm 0.13 \pm 0.03$ &  786.22   & $ 1.40 \pm 0.54 \pm 0.05$ & 1018.64   & $51.75 \pm 2.18 \pm 1.76$    \\
 769.41   & $ 0.40 \pm 0.17 \pm 0.04$ &  792.33   & $ 0.81 \pm 0.26 \pm 0.03$ & 1019.62   & $55.46 \pm 1.85 \pm 1.89$    \\
 778.19   & $ 1.10 \pm 0.51 \pm 0.04$ &  823.48   & $ 0.30 \pm 0.14 \pm 0.01$ & 1020.58   & $40.71 \pm 2.25 \pm 1.39$    \\
 780.12   & $ 1.27 \pm 0.52 \pm 0.04$ &  931.67   & $ 0.24 \pm 0.10 \pm 0.01$ & 1021.64   & $24.89 \pm 2.01 \pm 0.85$    \\
 781.17   & $ 1.37 \pm 0.50 \pm 0.05$ &  992.14   & $ 0.54 \pm 0.23 \pm 0.02$ & 1022.78   & $12.01 \pm 1.31 \pm 0.41$    \\
 782.06   & $ 1.70 \pm 0.46 \pm 0.06$ & 1009.68   & $ 3.30 \pm 1.04 \pm 0.11$ & 1027.76   & $ 2.86 \pm 0.48 \pm 0.10$    \\
 783.17   & $ 2.22 \pm 0.54 \pm 0.08$ & 1015.64   & $15.03 \pm 1.67 \pm 0.51$ & 1036.96   & $ 0.46 \pm 0.11 \pm 0.02$    \\
 784.27   & $ 1.97 \pm 0.55 \pm 0.07$ & 1016.70   & $22.64 \pm 1.81 \pm 0.77$ & 1055.64   & $ 0.094 \pm 0.034 \pm 0.007$ \\
 785.40   & $ 1.90 \pm 0.69 \pm 0.07$ & 1017.66   & $37.05 \pm 2.17 \pm 1.26$ &           &                              \\
\end{tabular}
\end{ruledtabular}
\end{table}

\begin{table}
\caption{\label{TabBrCompar} The $\rho,\omega,\phi \to \eta \gamma$ decay branching fractions.}
\begin{ruledtabular}
\begin{tabular}[t]{lccc} 
& Published~\cite{snd_etag} & This work & Previous measurements  \\ \hline
$B(\rho\to\eta\gamma)\cdot 10^4$   & $2.40 \pm 0.25 \pm 0.07$    & $2.82 \pm 0.30 \pm 0.17$    & $3.28 \pm 0.37 \pm 0.23$~\cite{cmdetagamma} \\
$B(\omega\to\eta\gamma)\cdot 10^4$ & $4.63 \pm 0.46 \pm 0.13$    & $4.33 \pm 0.44 \pm 0.17$    & $5.10 \pm 0.72 \pm 0.34$~\cite{cmdetagamma} \\
$B(\phi\to\eta\gamma)\cdot 10^2$   & $1.362 \pm 0.019 \pm 0.035$ & $1.364 \pm 0.023 \pm 0.044$ & $1.338 \pm 0.012 \pm 0.52$~\cite{SND_3g} \\
\end{tabular}
\end{ruledtabular}
\end{table}

\end{document}